\documentclass[aps,prl,twocolumn,showpacs,superscriptaddress,groupedaddress]{revtex4}  
\usepackage{graphicx}  
\usepackage{dcolumn}   
\usepackage{bm}        
\usepackage{amssymb}   
\usepackage[normalem]{ulem}

\usepackage{doi}
\usepackage{hyperref}

\begin{document}



\title{Symmetry of the Fermi surface and evolution of the electronic structure across the paramagnetic-helimagnetic transition in MnSi/Si(111)}

\author{Alessandro Nicolaou} 
\email{nicolaou@synchrotron-soleil.fr} 
\affiliation{Synchrotron SOLEIL, L'Orme des Merisiers, Saint-Aubin, BP 48, F-91192 Gif-sur-Yvette Cedex, France}

\author{Matteo Gatti}
\affiliation{Synchrotron SOLEIL, L'Orme des Merisiers, Saint-Aubin, BP 48, F-91192 Gif-sur-Yvette Cedex, France}%
\affiliation{Laboratoire des Solides Irradi\'{e}s, \'{E}cole Polytechnique, CNRS-CEA/DSM, F-91128 Palaiseau, France}
\affiliation{European Theoretical Spectroscopy Facility (ETSF)}

\author{Elena Magnano}
\affiliation{CNR-IOM, Laboratorio TASC, S.S. 14 Km 163.5, 34149 Basovizza, Trieste, Italy}%

\author{Patrick Le F\`{e}vre}
\affiliation{Synchrotron SOLEIL, L'Orme des Merisiers, Saint-Aubin, BP 48, F-91192 Gif-sur-Yvette Cedex, France}%

\author{Federica Bondino}
\affiliation{CNR-IOM, Laboratorio TASC, S.S. 14 Km 163.5, 34149 Basovizza, Trieste, Italy}%

\author{Fran\c{c}ois Bertran}
\affiliation{Synchrotron SOLEIL, L'Orme des Merisiers, Saint-Aubin, BP 48, F-91192 Gif-sur-Yvette Cedex, France}%

\author{Antonio Tejeda }
\affiliation{Synchrotron SOLEIL, L'Orme des Merisiers, Saint-Aubin, BP 48, F-91192 Gif-sur-Yvette Cedex, France}
\affiliation{Universit\'{e} de Lorraine, UMR CNRS 7198, Institut Jean Lamour, BP 239, F-54506 Vandoeuvre-lès-Nancy, France}%

\author{Mich\`{e}le Sauvage-Simkin}
\affiliation{Synchrotron SOLEIL, L'Orme des Merisiers, Saint-Aubin, BP 48, F-91192 Gif-sur-Yvette Cedex, France}%

\author{Alina Vlad}
\affiliation{Synchrotron SOLEIL, L'Orme des Merisiers, Saint-Aubin, BP 48, F-91192 Gif-sur-Yvette Cedex, France}%

\author{Yves Garreau}
\affiliation{Synchrotron SOLEIL, L'Orme des Merisiers, Saint-Aubin, BP 48, F-91192 Gif-sur-Yvette Cedex, France}%
\affiliation{MPQ, Universit\'{e} Denis Diderot Paris VII, B\^at. Condorcet, F-75205 Paris Cedex 13, France}%

\author{Alessandro Coati}
\affiliation{Synchrotron SOLEIL, L'Orme des Merisiers, Saint-Aubin, BP 48, F-91192 Gif-sur-Yvette Cedex, France}%

\author{Nicolas Gu\'{e}rin}
\affiliation{Department of Chemistry and Biochemistry, University of Bern, Freiestrasse 3, 3012 Bern, Switzerland}%

\author{Fulvio Parmigiani}
\affiliation{Elettra-Sincrotrone Trieste, Strada Statale 14, 34149 Basovizza, Trieste, Italy}
\affiliation{Universit\`{a} degli Studi di Trieste, Via A. Valerio 2, 34127 Trieste, Italy}
\affiliation{International Faculty - University of Cologne, Germany}

\author{Amina Taleb-Ibrahimi}
\affiliation{Synchrotron SOLEIL, L'Orme des Merisiers, Saint-Aubin, BP 48, F-91192 Gif-sur-Yvette Cedex, France}%
\date{\today}

\begin{abstract}
MnSi has been extensively studied for five decades, nonetheless detailed information on the Fermi surface (FS) symmetry is still lacking. This missed information prevented from a comprehensive understanding the nature of the magnetic interaction in this material. Here, by performing angle-resolved photoemission spectroscopy on high-quality MnSi films epitaxially grown on Si(111), we unveil the FS symmetry and the evolution of the electronic structure across the paramagnetic-helimagnetic transition at T$_C$ $\sim$ 40 K, along with the appearance of sharp quasiparticle emission below T$_C$. The shape of the resulting FS is found to fulfill robust nesting effects. These effects can be at the origin of strong magnetic fluctuations not accounted for by state-of-art quasiparticle self-consistent GW approximation. From this perspective, the unforeseen quasiparticle damping detected in the paramagnetic phase and relaxing only below T$_C$, along with the persistence of the d-bands splitting well above T$_C$, at odds with a simple Stoner model for itinerant magnetism, open the search for exotic magnetic interactions favored by FS nesting and affecting the quasiparticles lifetime. 
\end{abstract}

\maketitle

Transition-metal monosilicides, such as TM-Si (TM = Mn, Fe, Co) and their solid solutions, show an intriguing evolution of the electronic and magnetic properties as a function of temperature and doping \cite{Manyala6778_2000,Manyala1038_2008}. Among these materials, MnSi is a paradigmatic case for the exotic magnetic properties not accounted for by standard models of magnetism. At T$_C$ ($\sim$ 40 K in thin films) MnSi shows a phase transition from a paramagnetic metal state to a helimagnetic order. The effective magnetic moment $\mu_{eff}$  of 2.27$\mu_B$/Mn drops to $\mu_{sat}=$ 0.4$\mu_B$/Mn in the saturated ferromagnetic phase \cite{Wernick19721431,PhysRevB.76.052405}. This value is significantly smaller than the $1\mu_B$/Mn value predicted by local-density approximation (LDA) calculations \cite{Jeong}. Furthermore, the magnetic transition can be progressively suppressed by applying a relatively low hydrostatic pressure of 14.6 kbar, while T$_C$ tends to zero and an abrupt change of the resistivity from the T$^2$ Fermi liquid behavior to the T$^{3/2}$ non-Fermi liquid (NFL) character is observed. This NFL phase is quite robust with respect to temperature, pressure and magnetic fields, and it is accompanied by a partial magnetic order for p$<$21 kbar \cite{QCP,natureNonFermi,natureFermiBreakdown}. These properties are only partially accounted for by the Moriya's self-consistent renormalisation theory for spin fluctuations \cite{moriyabook}, hence the nature of the magnetic transition and the mechanism of the strong correlations in MnSi is matter of contests \cite{PRB_carbone,PRB_Mena}. In this complex and fascinating scenario MnSi has been extensively studied for the last five decades. Nonetheless, the unknown valence-band structure and FS symmetry have banned to unveil the nature of the magnetic interactions in this compound. The nesting of the Fermi surface (FS), as proposed by O. Narikiyo \cite{JPSJ.73.2910}, could account for some of the anomalies observed. This topological property of the FS could indeed create correlation effects accompanied by an important damping of the quasiparticles (QP). Moreover, the strong FS instabilities produced by a perfect nesting could induce low temperature phase transitions in which ordinary Fermi liquid behavior is recovered \cite{PhysRevB.42.4064}. Hence, the detailed knowledge of the FS topology is essential for a deeper understanding of the interplay between the complex magnetic phase diagram of MnSi and the NFL phases. Furthermore, a similar mechanism, in connection with the chiral character of the magnetic interactions, is expected to govern the properties of skyrmion lattices, discovered in this compound a few years ago \cite{Muhlbauer13022009}. Unfortunately, the lacking of a cleaving plane in MnSi single crystals has prevented, so far, momentum-resolved photoelectron spectroscopies, prompting us to epitaxially grow very high quality MnSi layers on Si(111). Details on sample preparation and data acquisition are reported in \cite{SM}. 

By performing angle-resolved photoemission spectroscopy (ARPES) experiments on 5.5 nm thick MnSi epitaxial films, we have been able to disclose the FS symmetry, revealing strong deviations of the experimental data from state-of-art band structure calculations as well as the potential for strong nesting for the most of the FS sheets, accompanied by significant QP broadening. Moreover, following the evolution of the electronic structure across the paramagnetic-helimagnetic phase transition at T$_C$ $\sim$ 40 K we detect, at some particular k$_F$ of the nested band, a notable sharpening of the quasiparticles. From this perspective, the unusual QP broadening of the paramagnetic phase which is relaxing only below T$_C$, along with the persistence of the d-bands splitting in the paramagnetic phase, in contrast with a simple Stoner model for itinerant magnetism \cite{PhysRevLett.88.167205,PhysRevLett.104.237204}, point to novel magnetic interactions not accounted for by state-of-art quasiparticle self-consistent GW approximation (QSGW) \cite{PhysRevLett.93.126406,PhysRevLett.96.226402}. Hence, our data are disclosing a possible major role of the FS nesting properties for developing strong magnetic fluctuations affecting the quasiparticles lifetime and thus expected to govern the magnetic interactions in this compound. 

	\begin{figure}
	
	\includegraphics[width=0.5\textwidth]{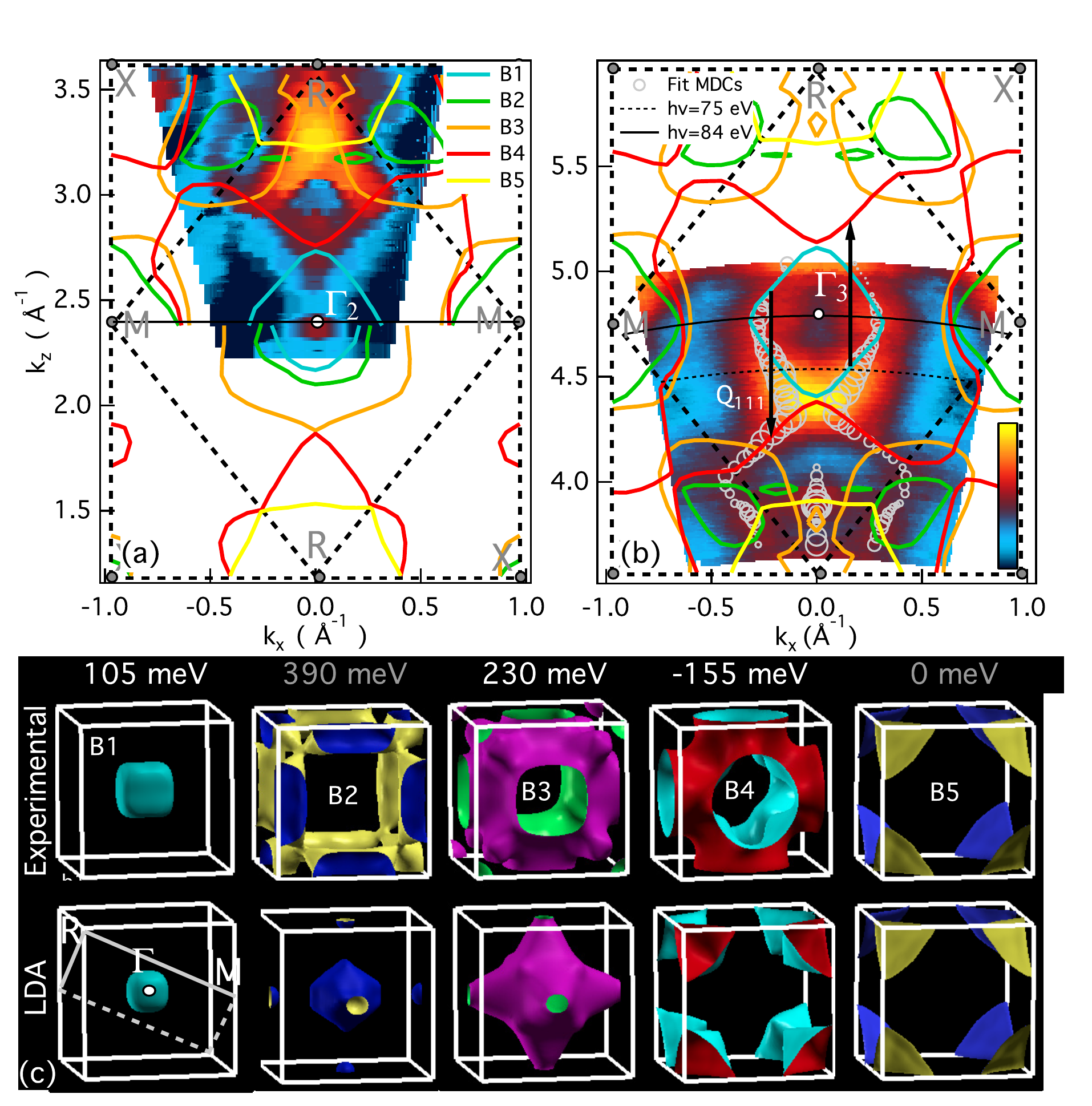}
	
			\caption{ \label{FS_GMR}  (a-b): Experimental FS mappings of the $\Gamma$MR plane done in normal emission and calculated using an inner potential V$_0 =$ 7 eV \cite{kzvsphoton}. (a) Intensity map at E$_{F}$ of the second derivative of the EDMs from a 15-45 eV energy scan in LH. (b) Intensity map at E$_{F}$ from a 50-98 eV energy scan in LV. The k$_F$ values, plotted as gray circles in (b), were obtained by the Lorentzian fit of the MDC lines at E$_F$. Black arrows are the nesting vectors. (c) FS as predicted by the LDA with the original position of  E$_{F}$ (bottom) and with E$_{F}$ shifted for the different bands in order to fit experimental data (top). Robust values in white, less compelling ones in gray, see text.}
			
	\end{figure}

\begin{figure}
	
	\includegraphics[width=0.5\textwidth]{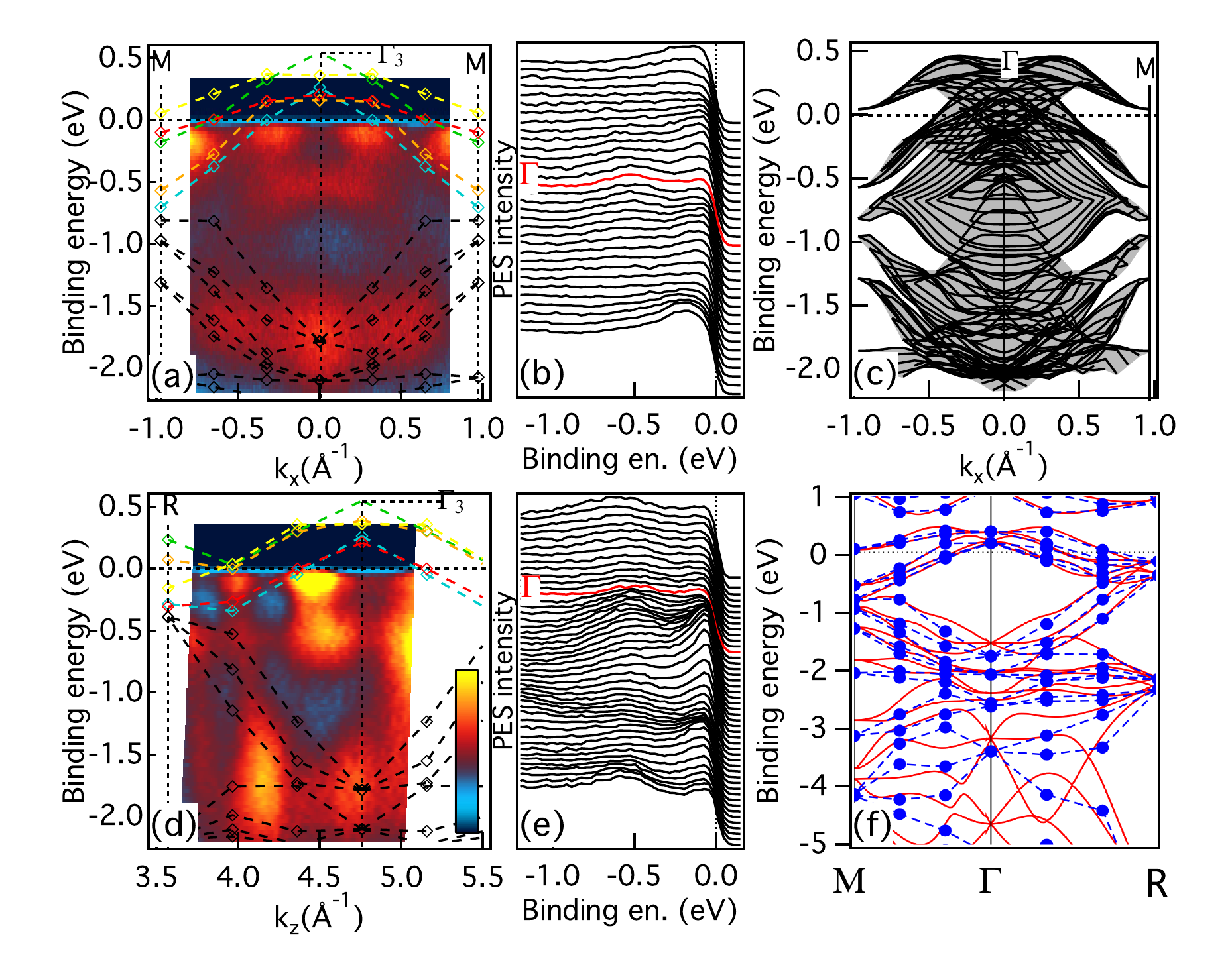}
	
			\caption{\label{DISPERSIONS} (a,d) EDM and (b,e) corresponding EDCs of the $\Gamma$M and $\Gamma$R directions, acquired at T=55 K using LV polarization. The superimposed QSGW bands are shifted in energy to match the experimental FS of Fig. \ref{FS_GMR}(a,b). (c) Projection of the LDA bands of the $\Gamma$MR plane along $\Gamma$M. (f) Band-structure calculations in LDA (red solid lines) and QSGW (blue markers; linearly interpolated dotted lines are a guide for the eyes).}
			
	\end{figure}

Figures \ref{FS_GMR}(a-b) show FS mappings of the $\Gamma$MR plane in the 2.17 to 5.16 \AA$^{-1}$ k$_z$ range, across the second and third Brillouin zones (BZ), centered at $\Gamma_{2}$=(0,0,2.38) and $\Gamma_{3}$=(0,0,4.76) \AA$^{-1}$ \cite{kzvsphoton}. Both measurements were performed in the paramagnetic phase on two different samples of the same thickness, but with different photon polarization (linear horizontal LH and vertical LV). The FS in (a) was measured at 300 K in LH. The FS in (b) was obtained in LV at lower temperature (55 K) for reducing the thermal broadening of the quasiparticles. LDA and QSGW calculations results give the same description of the five $3d$ bands crossing E$_{F}$, allowing to use the computationally cheaper LDA for the FS analysis, however QSGW noticeably improve the agreement with the experimental data for the Mn $3d$ - Si $sp$ hybridized states at higher binding energies [see Fig. \ref{DISPERSIONS}(f)]. In the bottom half of Fig. \ref{FS_GMR} (a) we plot the original LDA contours. In the top half of Fig. \ref{FS_GMR} (a), and in (b), we present the best match to our data, which was obtained by displacing each calculated band (B1 to B5) with respect to E$_F$ by the amount indicated in the top row of Fig. \ref{FS_GMR} (c). We note that only by operating such a charge redistribution within the 3$d$ manifold we can properly describe the experimental data. Actually, while the LDA predicts three hole-like pockets centered at $\Gamma$, and two electron-like ones at R, we find instead that the biggest hole-pocket B3 is forming long tubes containing the R points and that the biggest electron-pocket B4 has increased in size. Contrarily to B3, B4 is clearly visible also in the third BZ in LV polarization, where its FS contour and the one of B1, recognized by its rhomboidal shape, were tracked more precisely by a Lorentzian fit of the momentum distribution curves (MDCs) [gray circles in Fig. \ref{FS_GMR}(b)]. This latter measurement reveals indeed that the hole-like pocket B1, centered at $\Gamma_{3}$ and not visible in the second BZ and/or with LH polarization, has also increased in size. The certain identification of the B1,B3,B4 bands is provided by their characteristic shape together with the very good agreement with the intensity maps of Fig. \ref{FS_GMR}(a-b), making the corresponding energy shifts of 105, 230 and -155 meV, quite robust values. Less compelling are the values proposed for B2, lifted up by 390 meV to better account for the intensity parallel to the MR directions, and for B5, not shifted, as the present data do not allow to unambiguously identify them.

In a similar way, as cited by Carbone in \cite{CarbonePhD}, S. B. Dougale and J. Laverock imposed a common rigid shift of 200 meV to the calculated bands in order to describe their positron annihilation measurements. However, this approach does not conserve the number of particles. Conversely, to describe our data we have lowered E$_{F}$ only for the three LDA bands B1,B2,B3 while we have increased it for B4, in a way that might conserve the total number of particles. Even if  a precise determination of the electron counting is meaningless without a complete 3D FS mapping, the observed disagreement with theory is unlikely originated from an uncontrolled doping of the sample and/or surface polarity effects, that would produce the same shift of E$_F$ for all bands, whereas it calls for other explanations.

Interestingly, following this unforeseen charge redistribution, the experimental FS results more prone to nesting phenomena. The B1 FS sheet has indeed acquired a rhomboidal shape with parallel sections connected by nesting vectors 0.53 \AA$^{-1}$ long, lying parallel to the [201] and [021] directions. Moreover, the top (bottom) half of B1 and the B4 band at lower (higher) k$_z$ are almost perfectly nested, creating a coupling between hole- and electron-pockets with wave vectors 0.73 \AA$^{-1}$ long. It is remarkable that this latter lies parallel to the [111] direction, along which the direction of the helimagnetic order is pinned below T$_C$, and that, starting at about the same wave vector, neutron scattering experiments detect a strong temperature dependence of the magnetic excitations, not predicted by the Moriya's spin-fluctuations theory \cite{CarbonePhD}. Moreover, FS nesting is expected to produce unusual variations of the optical reflectivity through the enhancement of QP scattering \cite{PhysRevB.43.5498}, being thus a plausible responsible for the non-Drude optical behavior of MnSi reported in \cite{PRB_Mena}.

\begin{figure}

		\includegraphics[width=0.5\textwidth]{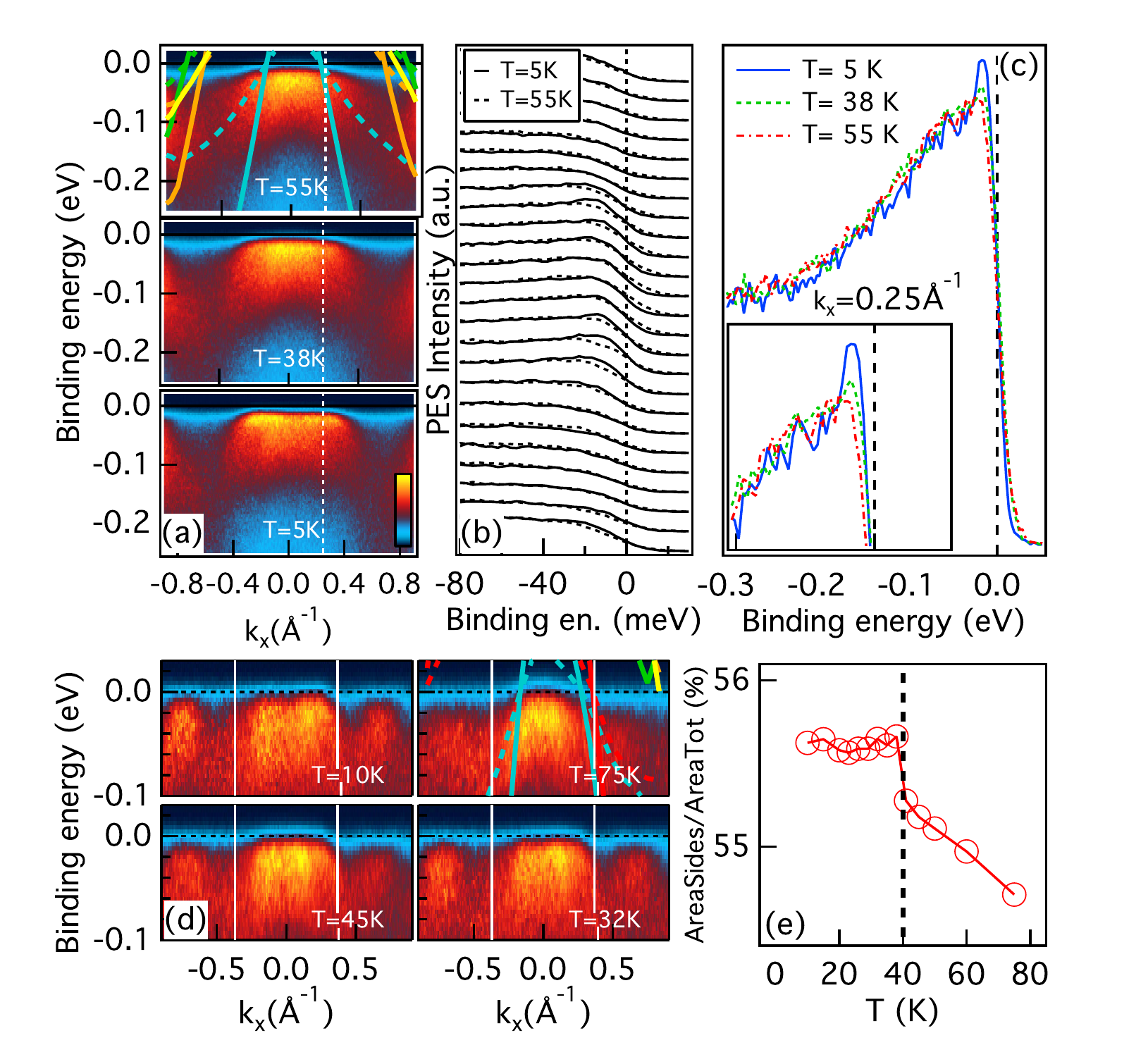}

			\caption{ \label{T_dep}  (a) EDM images as a function of temperature acquired in LV with photons of 75 eV. They correspond to a cut of the reciprocal space parallel to $\Gamma$M but at $k_{z}=4.5$ \AA$^{-1}$, see dotted line in Fig. \ref{FS_GMR}(b). (b) Comparison of the highest (dotted) and lowest (thick) temperatures EDCs of the data presented in (a). (c) EDCs as a function of temperature at k$_x=0.25$ \AA$^{-1}$. (d) Temperature dependence of EDM images taken at the same $k_{z}$ as (a) but rotated by 30 $^{\circ}$ with respect to $\Gamma$M. (e) Fraction of the spectral intensity at the sides ($k_{x}\geq$ 0.38 \AA$^{-1}$) and between -0.1 eV and E$_{F}$ as a function of temperature. In (a) and (c) solid lines represents LDA bands, and the dotted one LDA bands renormalized by 3.3.}

	\end{figure}

The indication that we are in presence of a correlated metal arises from the strong damping of the quasiparticles observed along both the $\Gamma$M and the $\Gamma$R high-symmetry directions, as presented in Fig. \ref{DISPERSIONS}(a,b) and (d,e) respectively. On the top of the energy dispersion maps (EDMs) we plot the result of QSGW calculations, where the Mn $3d$ bands are shifted in energy according to the values used to reproduce our experimental FS. If the Fermi level crossings are clearly discernible from the FS maps, below E$_{F}$ the lineshape becomes very broad and, accordingly, a significant mass renormalization is clearly visible in the top panel of Fig. \ref{T_dep}(a), where the dotted line represents the B1 LDA band renormalized by a factor 3.3, a value which is consistent with previous Haas van-Alphen and infrared spectroscopy experiments \cite{Taillefer1986957,PRB_Mena}. We note that while the LDA eigenvalues in principle cannot be interpreted as the electron addition and removal energies which are measured by ARPES, the renormalization and the rigid shift of the d-bands binding energies with respect to those calculated by the many-body GW approximation (GWA) \cite{PhysRev.139.A796}, where the self-energy is given by a product of the one-particle Green's function $G$ and the screened Coulomb interaction $W$ and therefore contains electron-hole and plasmon excitations, points to exotic magnetic interactions. Actually, the resulting dynamical screening of the photoemission hole is a crucial ingredient if we want to compare ARPES with band structure calculations, however magnetic correlation effects due to the coupling with spin excitations are not accounted for by the GWA  \cite{PhysRevLett.80.2389,PhysRevLett.93.096401,PhysRevB.73.125105,PhysRevB.85.155131,qsgw}. Therefore, we conclude that this large broadening is intrinsic to the electronic excitations of MnSi and it is the fingerprint of a many-body mechanism associated with the spin degree of freedom.

Even if part of this broadening arises from the superposition of the FS sheets corresponding to MnSi(111) left- and right- handed twinned domains revealed by our GIXD measurements \cite{SM}, this effect is expected to be smaller with respect to the observed one. Moreover, several factors, like final-state effects for 3D electronic structures \cite {Strocov200365} or corrections to the sudden approximation \cite{PhysRevB.58.15565}, can make the measured photocurrent significantly differ from the intrinsic spectral function of a material. Nevertheless, those effects are expected to be strongly photon-energy dependent, while we do not observe any significant variation in 15-100 eV energy range. In addition, we detect another spectral feature centered at $\Gamma$ with the top at about 500 meV below E$_F$ and not predicted by GW spectral-function calculations \cite{SM}. Even though the dispersion of this feature in the $\Gamma$R direction is modest, it appears in a position forbidden for pure surface states, as it is already occupied by bulk bands, projected in Fig. \ref{DISPERSIONS}(c) on the $\Gamma$M direction. Moreover, it does not fit the periodicity of the $\bar{\Gamma}$\={M} surface direction, which is half of $\Gamma$M periodicity. The origin of this spectral feature, either a real hole-like band sunk below E$_F$ or the incoherent component associated with one of the Mn $3d$ bands at E$_{F}$, calls for a deeper investigation which is beyond the scope of the present work.

The effect of the magnetic transition on the electronic structure B1 is presented in Fig. \ref{T_dep}. In (a) we show a cut of the reciprocal space parallel to the $\Gamma$M direction but at a lower k$_z$ [see dotted line in Fig. \ref{FS_GMR}(b)]. Below 40 K we observe the appearance of sharp quasiparticle emission at about k$_x$=0.25 \AA$^{-1}$, see Fig. \ref{T_dep}(c). This sharp feature proves that the intrinsic limitation of the quasiparticle lifetime in the paramagnetic phase is relaxed when the system, cooled down below T$_C$, enters the ordered helimagnetic phase. In Fig. \ref{T_dep}(b) we compare the energy dispersion curves (EDCs) for the highest and lowest temperatures measured here, to show that the rising of the sharp feature below T$_C$ unambiguously originates from an increased quasiparticle lifetime and not by a mere increase of energy resolution at low T. To complete the interpretation of the magnetic transition we have precisely tracked the evolution of the electronic structure across T$_C$. To this end, temperature-dependent cuts of the reciprocal space at the same k$_z$, but rotated 30$^\circ$ with respect to the $\Gamma$M direction, have been considered. Fig. \ref{T_dep}(d) reports such data. By decreasing the temperature we detect the appearance of two symmetric bands at about $\pm$0.75 \AA$^{-1}$. The evolution of these features through the magnetic transition has been tracked by following their relative spectral weight with respect to the whole EDM in Fig. \ref{T_dep}(e). They have a constant weight below T$_C$, while at T$_C$ they start to loose progressively weight until 75 K, which is the maximum temperature measured here. It is known that the magnetic transition is suppressed in thin films by reducing the film thickness \cite{PhysRevB.82.184417,PhysRevB.88.115433}. The observed T$_C$ and the thickness of our film measured by SQUID and GIXD respectively \cite{SM}, are consistent with these studies, thus endorsing our novel recipe for growing MnSi on Si(111). Despite the fact that the spectral weight evolution of these bands allows to track correctly the temperature dependence of the magnetic transitions, they do not correspond to the calculated electronic bands and their identification calls for further investigations beyond the scope of the present work. Nevertheless, the persistence of an effect of magnetism well above T$_{C}$ implies the presence of local magnetic order well above T$_{C}$ \cite{PhysRevLett.88.167205,PhysRevLett.104.237204}.

In conclusion, by performing ARPES experiments on high-quality MnSi epitaxial layers, we have disclosed the FS symmetry along with robust nesting effects that can be at the origin of strong electron-electron and electron-hole scattering, not anticipated by the state of the art of band structure calculations. In this scenario, the unusual quasiparticles damping detected in the paramagnetic phase and relaxing only below T$_{C}$, along with the persistence of the d-bands splitting in the paramagnetic phase, in contrast with standard models of itinerant magnetism, point to novel magnetic interactions not accounted for by quasiparticle self-consistent GW approximation. Interestingly, this scenario seems to gain support by following the evolution of the electronic structure across the paramagnetic-helimagnetic transition at T$_{C}$ $\sim$ 40 K, where we observe an unforeseen sharpening of the quasiparticles at some particular k$_F$ values of the nested band. Henceforth, we propose here that the nesting properties of the MnSi Fermi surface, by affecting the quasiparticles lifetime through the development of strong magnetic fluctuations, govern the magnetic interactions present in this compound.

The authors thank K. Hricovini, P. de Padova, V. Brouet, F. Sirotti and S. Populoh for useful discussions.
This research was supported by a Marie Curie FP7 Integration Grant within the 7th European Union Framework Programme.
Computational time was granted by GENCI (Project No. 544).   

\bibliography{Biblio_corrected.bib}













\end{document}




\title{Symmetry of the Fermi surface and evolution of the electronic structure across the paramagnetic-helimagnetic transition in MnSi/Si(111) 

Supplementary Material}

\author{Alessandro Nicolaou}
\email{nicolaou@synchrotron-soleil.fr} 
\affiliation{Synchrotron SOLEIL, L'Orme des Merisiers, Saint-Aubin, BP 48, F-91192 Gif-sur-Yvette Cedex, France}

\author{Matteo Gatti}
\affiliation{Synchrotron SOLEIL, L'Orme des Merisiers, Saint-Aubin, BP 48, F-91192 Gif-sur-Yvette Cedex, France}%
\affiliation{Laboratoire des Solides Irradi\'{e}s, \'{E}cole Polytechnique, CNRS-CEA/DSM, F-91128 Palaiseau, France}
\affiliation{European Theoretical Spectroscopy Facility (ETSF)}

\author{Elena Magnano}
\affiliation{CNR-IOM, Laboratorio TASC, S.S. 14 Km 163.5, 34149 Basovizza, Trieste, Italy}%

\author{Patrick Le F\`{e}vre}
\affiliation{Synchrotron SOLEIL, L'Orme des Merisiers, Saint-Aubin, BP 48, F-91192 Gif-sur-Yvette Cedex, France}%

\author{Federica Bondino}
\affiliation{CNR-IOM, Laboratorio TASC, S.S. 14 Km 163.5, 34149 Basovizza, Trieste, Italy}%

\author{Fran\c{c}ois Bertran}
\affiliation{Synchrotron SOLEIL, L'Orme des Merisiers, Saint-Aubin, BP 48, F-91192 Gif-sur-Yvette Cedex, France}%

\author{Antonio Tejeda }
\affiliation{Synchrotron SOLEIL, L'Orme des Merisiers, Saint-Aubin, BP 48, F-91192 Gif-sur-Yvette Cedex, France}%
\affiliation{Universit\'{e} de Lorraine, UMR CNRS 7198, Institut Jean Lamour, BP 239, F-54506 Vandoeuvre-lès-Nancy, France}%

\author{Mich\`{e}le Sauvage-Simkin}
\affiliation{Synchrotron SOLEIL, L'Orme des Merisiers, Saint-Aubin, BP 48, F-91192 Gif-sur-Yvette Cedex, France}%

\author{Alina Vlad}
\affiliation{Synchrotron SOLEIL, L'Orme des Merisiers, Saint-Aubin, BP 48, F-91192 Gif-sur-Yvette Cedex, France}%

\author{Yves Garreau}
\affiliation{Synchrotron SOLEIL, L'Orme des Merisiers, Saint-Aubin, BP 48, F-91192 Gif-sur-Yvette Cedex, France}%
\affiliation{MPQ, Universit\'{e} Denis Diderot Paris VII, B\^at. Condorcet, F-75205 Paris Cedex 13, France}%

\author{Alessandro Coati}
\affiliation{Synchrotron SOLEIL, L'Orme des Merisiers, Saint-Aubin, BP 48, F-91192 Gif-sur-Yvette Cedex, France}%

\author{Nicolas Gu\'{e}rin}
\affiliation{Department of Chemistry and Biochemistry, University of Bern, Freiestrasse 3, 3012 Bern, Switzerland}%

\author{Fulvio Parmigiani}
\affiliation{Elettra-Sincrotrone Trieste, Strada Statale 14, 34149 Basovizza, Trieste, Italy}
\affiliation{Universit\`{a} degli Studi di Trieste, Via A. Valerio 2, 34127 Trieste, Italy}
\affiliation{International Faculty - University of Cologne, Germany}%

\author{Amina Taleb-Ibrahimi}
\affiliation{Synchrotron SOLEIL, L'Orme des Merisiers, Saint-Aubin, BP 48, F-91192 Gif-sur-Yvette Cedex, France}%
\date{\today}

\pacs{79.60.-i,71.20.Lp,75.50.Cc,71.27.+a}
\maketitle

\section{\label{sec:level2}Data acquisition}

ARPES and GIXD measurements were performed at the beamlines CASSIOPEE and SIXS respectively, at the synchrotron SOLEIL. ARPES data were acquired using a Scienta R4000 hemispherical analyzer with the entrance slit orthogonal with respect to the plane containing the incident beam and the normal to the sample surface. The energy resolution for ARPES was set to 25 (35) meV and 10 meV for the FS mapping in linear vertical LV polarization (linear horizontal LH) and the temperature dependences respectively. GIXD measurements were done using an UHV z-axis diffractometer at 18.41 KeV.

\section{\label{sec:level2}ARPES data analysis}

In the presented ARPES data, each energy dispersion map (EDM) was normalized so that each EDC has the same area. The different EDMs where normalized one with respect to the other to have a continuous evolution of the intensity along the $\Gamma$R direction. This normalization procedure was used in order to reduce the effect of the matrix elements relative to the different final states for different the photon energies, and to highlight the symmetry of the FS surface in Fig. 1(b) and the dispersion of the electronic bands in Fig. 2 (d-e) of the main text.

\section{\label{sec:level1}State of the art of ARPES data on MnSi}

Grytzelius $et$ $al.$ report the electronic structure of a very thin (5-8  $\AA$) MnSi film on Si (111). They interpret their result as the electronic structure of the Si(111) surface in presence of MnSi terminations \cite{PhysRevB.78.155406}. Kura $et$ $al.$ report ARPES measurements of an MnSi(100) single crystal polished and annealed; they observe some band dispersion in the ligand Mn 3d Si sp states at high binding energy, but they do not resolve the Mn 3d bands and they do not observe any clear dispersion near E$_{F}$ \cite{ARPES_MnSi(100)}. Even higher energy resolved measurements of J.Y. Son $et$ $al.$ reported \cite{JPSJ.71.1728} about a scraped MnSi single crystal, do not show considerable variation of the band structure with temperature and look like the ones obtained on an oxidized sample. We stress the fact that one of the main limitation of the ARPES technique is its surface sensitivity (limited to the very first atomic layers in the UV regime) and the necessity of an atomically flat surface, as the measure of the k$_{//}$, the component of the wavevector $\textbf{k}$ parallel to the sample surface, is done by measuring the take-off angle with respect to the sample surface. Therefore, surface preparation of single crystals might change the stoichiometry of the first atomic layers, preventing a measure representative of the bulk and, likely, the measure of the magnetic transition which is strongly doping dependent in this compound. This might be the reason why, so far, no evolution of the electronic structure as a function of temperature across T$_C$ has been observed.

\section{\label{sec:level3}Details on sample synthesis and characterization}

The epitaxial MnSi films were prepared by simultaneous co-evaporation from Knudsen cell crucibles of Mn and Si on a Si(111)-7x7 substrate at room temperature, and subsequent annealing at 250 $^\circ$C for several hours. With respect to our previous works \cite{Magnano20063932,Carleschi20074066}, the long annealing time at a relatively low temperature of 250 $^\circ$C was found to be the key to obtain homogeneous films showing a well-defined band dispersion. The absence of spurious phases is proved by the grazing incidence X-ray diffraction (GIXD) and the presence of the magnetic transition by SQUID \cite{SM}.

\begin{figure}[htbp]

		\includegraphics[width=0.5\textwidth]{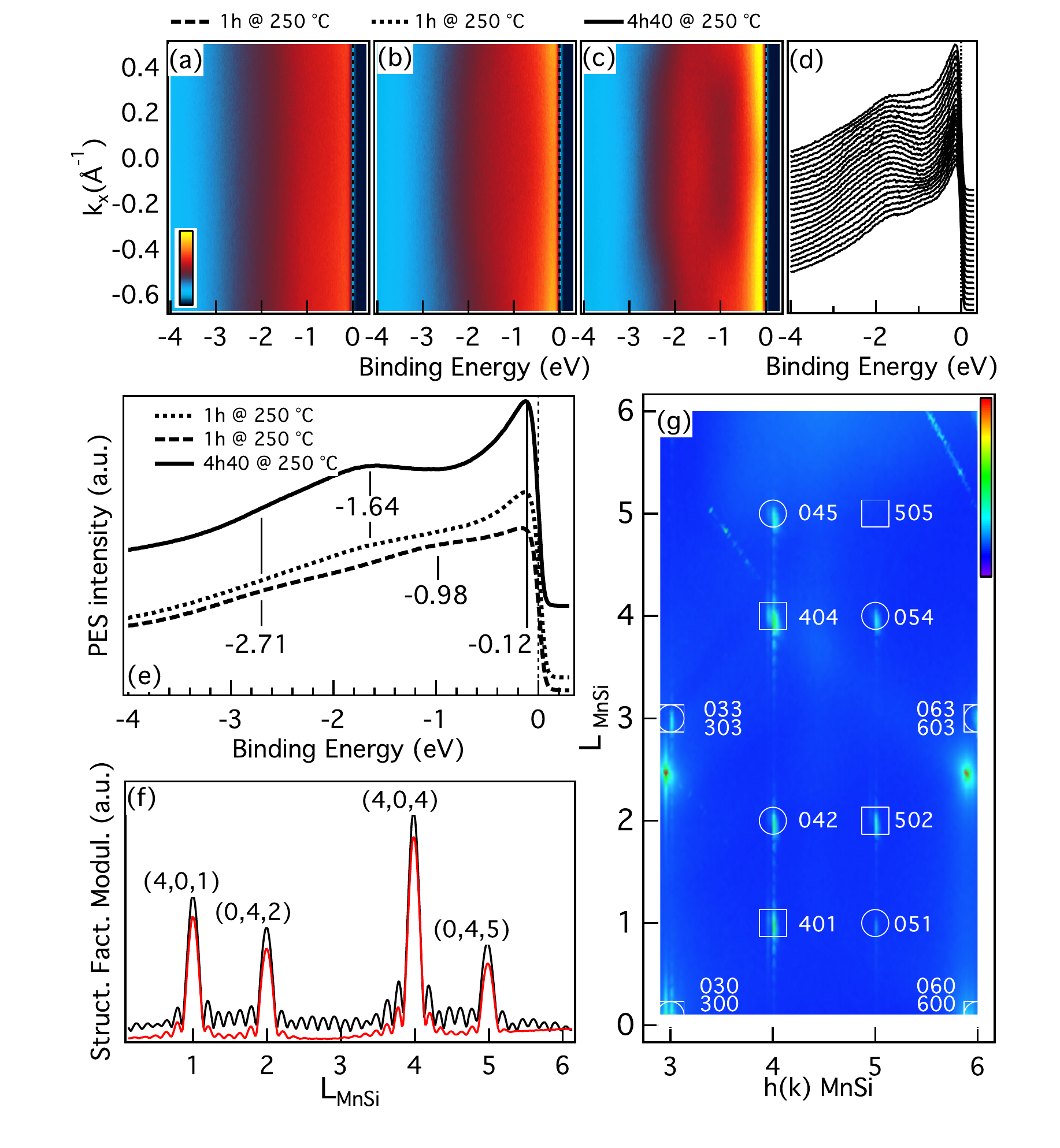}

			\caption{\label{GIXD} (a-b) ARPES data after 1h heating at 250 $^\circ$C in two different position on the sample. (c-d) ARPES data after 4h40min heating on an homogeneous sample and correspondent stacks. (e) Photoemission data obtained by integration of EDMs in (a-c) over the whole k-range. All these data where taken at room temperature using a 30 eV LH polarized beam. (f) l ROD at h(k)=4 compared to the fit. Data and fit where vertically shifted for clarity. (g) h(k),l map of 5.5 nm MnSi(111)/Si(111). GIXD measurements where performed at room temperature and ambient pressure on Si capped MnSi/Si(111) sample. h,k,l are expressed in the MnSi(111) basis. The bright spots not highlighted are the ones of the Si(111) substrate.}
	\end{figure}
 
In figure \ref{GIXD}(a-c) we show how the heating time strongly influences the homogeneity and crystallinity of our films.  Higashi $et$ $al.$\cite{Higashi} studied the temperature dependence of the LEED spot intensity and found that the best heating temperature is 250 $^\circ$C, corresponding to the maximum spot intensity, and therefore a well defined long-range order. In their work, they heated for 30 min 3 monolayer of Mn evaporated on the Si(111) substrate. Looking at our ARPES data, we have found that in order to obtain homogeneous samples showing a well defined band-dispersion it is necessary to heat at this relatively low temperature for a long time. The data presented in Fig. \ref{GIXD}(a,b) where both taken after 1 hour of heating at 250 $^\circ$C, but in different parts of the sample. As is clear from these two data sets, after only 1 h at 250 $^\circ$C the sample is not homogeneous and presents weakly dispersing bands only in some regions. Additional heating of the same sample for about 4 hours, yielded the nice dispersing bands presented in \ref{GIXD}(c,d), which were found to be homogeneous over the whole sample. The resulting dispersion of these bands is clear by looking at the corresponding Energy Dispersion Curves (EDC) presented in  \ref{GIXD}(d). If we look at the comparison of the integrated photoemission data presented in  \ref{GIXD}(e), obtained by integrating each energy dispersion map (EDM) over the whole k-range, we observe that, despite all of them present clear weight at the Fermi level (E$_F$), only in the long heated sample the Mn 3d states (max $@$ -0.12 eV) are well separated with respect to the bonding Mn 3$d$ hybridized with the Si $sp$ bands, which are found at -1.64 eV.

The homogeneity of the long heated sample and the presence of the single B20 crystalline phase is confirmed by the GIXD measurements presented in Fig. \ref{GIXD}(f-g). The diffraction pattern obtained by GIXD shows a very good epitaxy of the film with the crystallographic relationship with the substrate being MnSi(111)/Si(111). From reciprocal space maps parallel to the surface plane (not shown), we have identified the surface cells of the substrate and of the film, and we have measured a mismatch between the film and the substrate of 2.2$\%$, meaning that our film is not fully relaxed at its own lattice parameter, but 1.1$\%$ tensile strain is still present \cite{Suto2009226}. We recall that a fully relaxed MnSi(111) film on Si(111) present a 3.3$\%$ difference in the in-plane lattice parameter with respect to the substrate. From the the h(k),l map of the reciprocal plane in the base of MnSi presented in Fig.\ref{GIXD} (g) we have determined the the presence of 2 twinned domains. The two different sets of diffraction spots, expected from each B20 structure, are presented as circles and squares for the two domains. The presence of twinned domains is due to the growing of a non-centrosymmetric structure as the B20 one on a centrosymmetric substrate (Si(111)) \cite{Suto2009226}. Karhu et al. reported domains of few $\mu^2$ size \cite{PhysRevB.84.060404}. From a fit of our hk scans with a gaussian line shape we can estimate a minimum size of individual twin domains in our sample to less than 0.01 $\mu^2$. ARPES measurements were performed with a photon beam of about 100x100 $\mu^2$ size; we therefore illuminate both domains and we expect the mere superposition of the two band structures. By fitting the rod at h=k=4 presented in \ref{GIXD} (g) with ROD we can estimate the presence of 45 $\%$ of one domain and 55 $\%$ of the other. We can also estimate precisely the sample thickness to 55 $\AA$. The best fit was obtained by stacking 7 unit cells of MnSi and leaving free the c-strain of each unit cell, yielding a strain variation from -1.75 $\%$ at the interface to 0.5 $\%$ at the surface. By performing LDA calculation using a trigonally stressed B20 unit cell, both in and out the $<$111$>$ plane, we have observed only negligible variations of the electronic structure.

\begin{figure}[htbp]

		\includegraphics[width=0.5\textwidth]{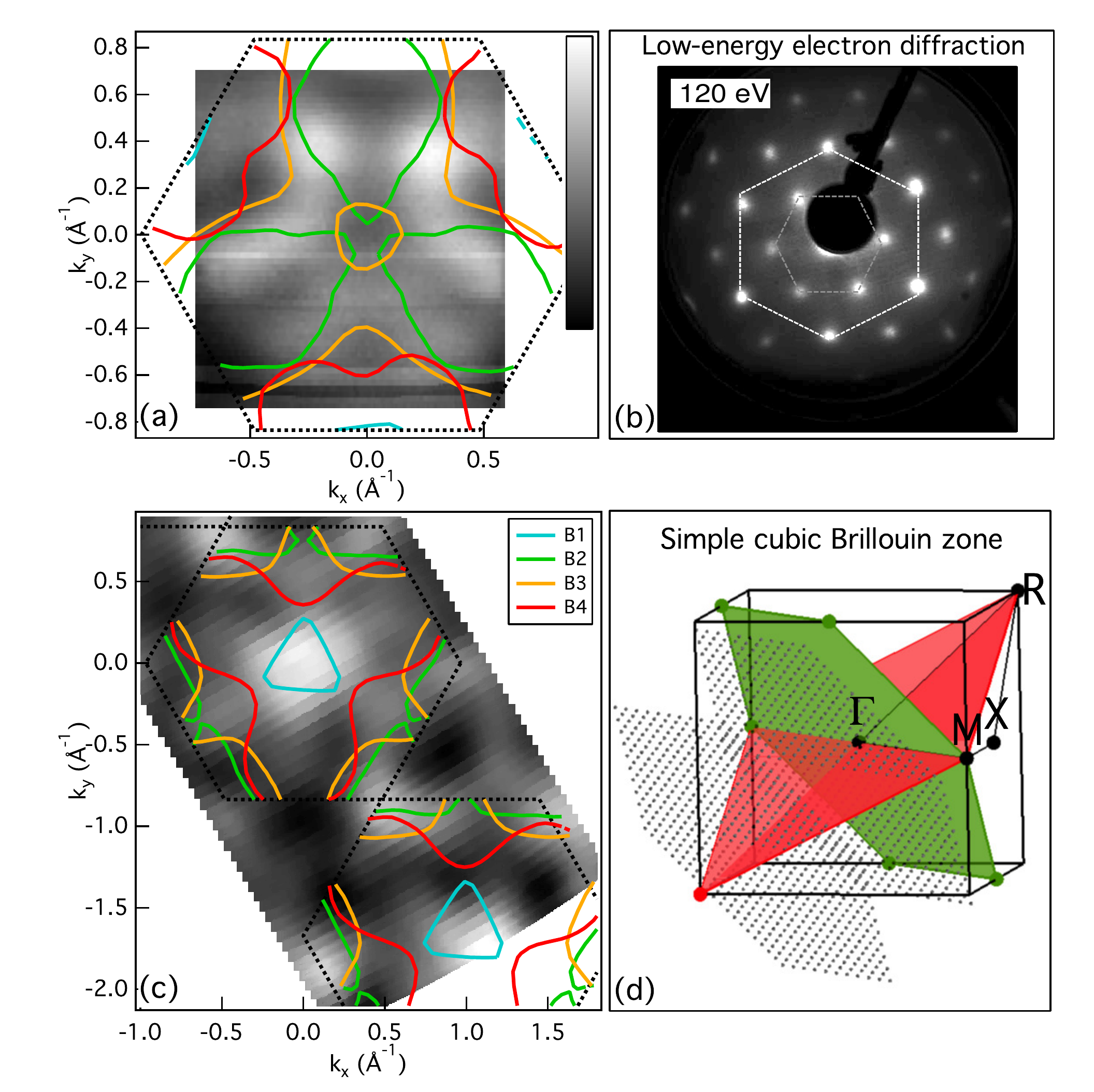}

			\caption{\label{LEED} (In (a) and (c) we present Fermi surfaces mapping acquired by performing polar scans at photon energies of 35 eV and 75 eV respectively. These are compared to LDA band structure calculations where the electronic bands were shifted in energy to reproduce the experimental Fermi surfaces presented in Fig. 1 of the main text. These two measurements probe planes of the reciprocal space parallel to the $\Gamma$MM one but shifted along the 111 direction of -16 $\%$  and - 67 $\%$ of the $\Gamma$R direction. The meshes used for the LDA calculation are presented in (d) together with the simple cubic Brillouin zone of the B20 crystal structure and the $\Gamma$MM plane (green) and the $\Gamma$MR one (red). In (b) we show the Low energy electron diffraction pattern of the MnSi thick film.}
	\end{figure}

In Fig. \ref{LEED} (a) and (c) we present Fermi surface (FS) mappings acquired at fixed photon energies of 35 eV and 75 eV using linear vertical and linear horizontal polarization respectively. These polar scans are compared with LDA band-structure calculations performed using an hexagonal mesh of the same size of the $\Gamma$MM plane and parallel to it, but shifted along the 111 direction of -16 $\%$  and - 67 $\%$ of the $\Gamma$R direction length. The calculated bands were shifted in energy by the same amount used to reproduce the experimental FS obtained by the energy dependencies and presented in the Fig. 1 of the main text. The good agreement between data and this simple model also away from the $\Gamma$MR plane further confirms the need to readjust the energy of the bands calculated by LDA in order to describe the experimental Fermi surface of this compound. The low energy diffraction pattern (LEED) is presented in (b), to show the long range order of the surface and no presence of surface reconstructions. The small hexagon correspond to the surface termination of the MnSi film, while the bigger one to the MnSi bulk. The sixfold symmetry of this latter is likely due to the presence of twinned domains.

\begin{figure}[hb]
\includegraphics[width=0.55\textwidth]{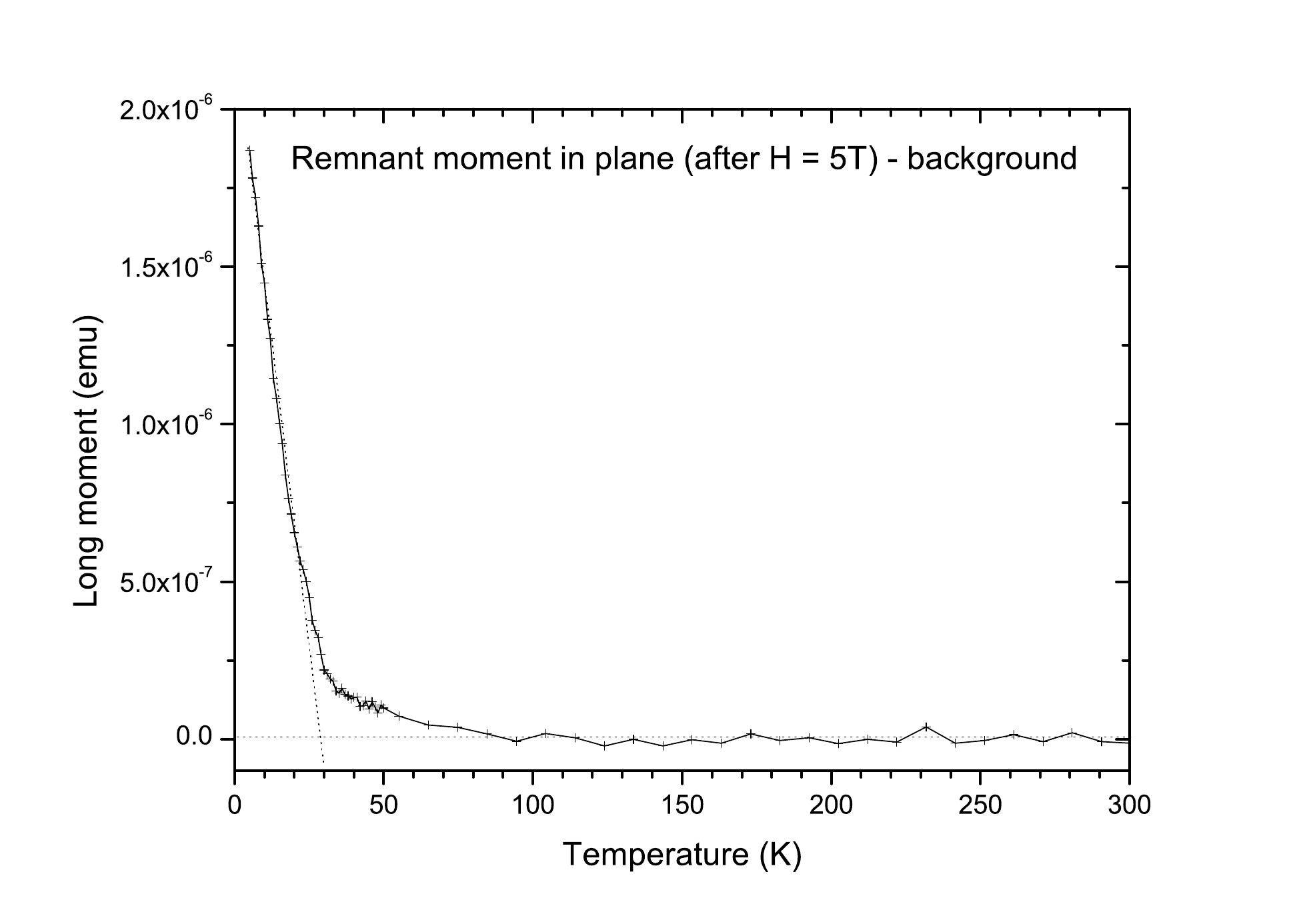}\\
\includegraphics[width=0.47\textwidth]{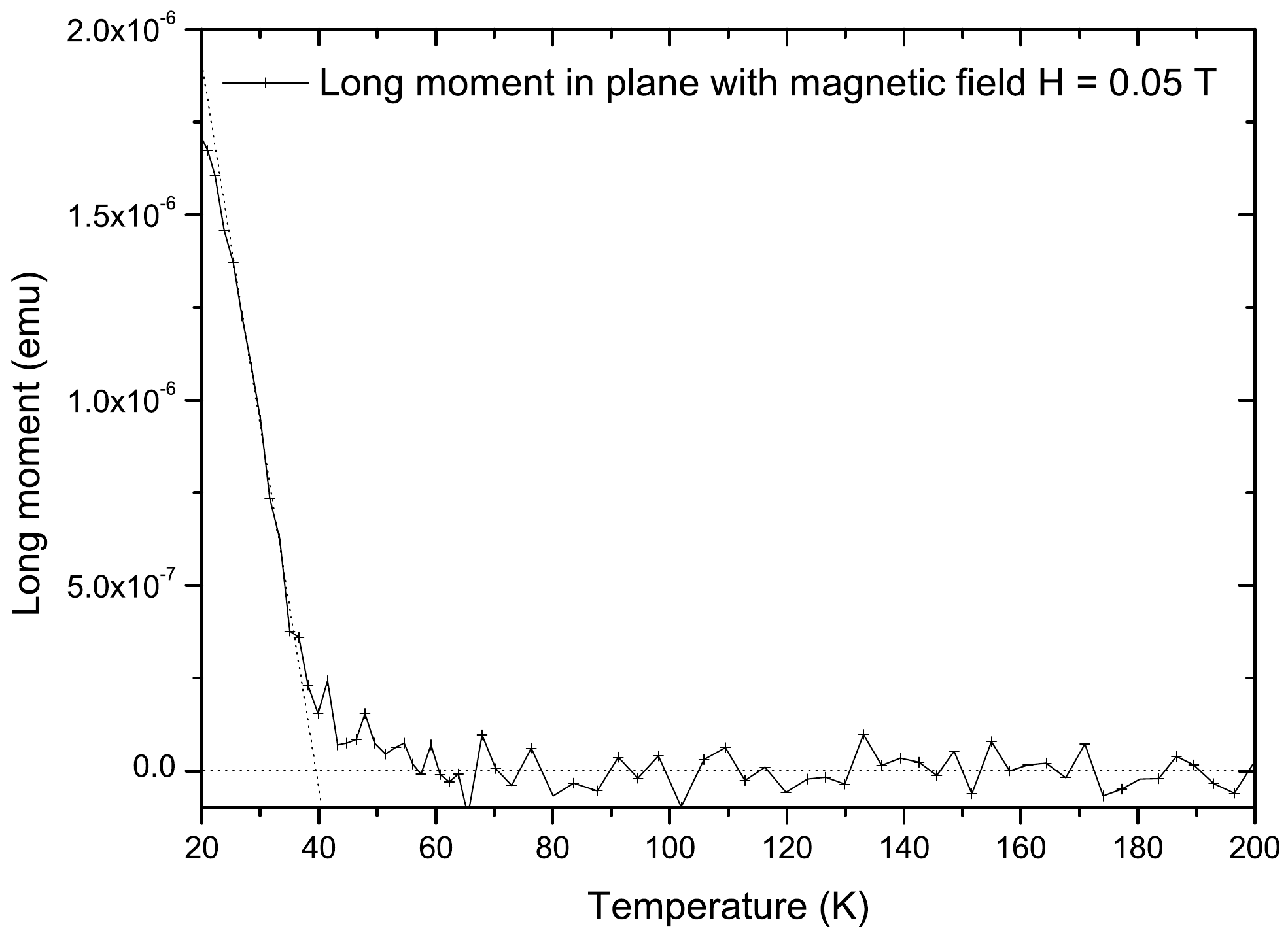}
\caption{\label{SQUID}In-plane magnetic moment under the field H = 0.05T for the thin film MnSi(111). Zero field cooling and subtraction of the Si(111) background moment}
\end{figure}

The magnetic transition of our 5.5 nm thick MnSi/Si(111) sample was characterized by means of SQUID magnetometer (MPMS Quantum Design), as presented in Fig. \ref{SQUID}. The MnSi thin film plane was set parallel to the applied magnetic field H so that the in-plane long moment is recorded. In a first strategy, in Fig. \ref{SQUID} (a), after a zero field cooling, H = 0.05 T was set and the moment was then measured for increasing temperature. In this way we observe a transition at T$_C$ = 40 K, after the subtraction of the Si(111) background. In a second strategy, in Fig. \ref{SQUID} (b), the sample was cooled down to 5 K and saturated by applying the field H = 5 T. After removing this field, the remnant moment was measured upon warming, yielding again a T$_C$ of 40 K. For all curves, we observe a continuous and straight decrease starting from 5 K down to T$_C$, which is compatible with weak itinerant ferromagnetism Above T$_C$, the background values are still positive even when the applied field was totally cancelled. Thus the sample may show an additional ferromagnetic component (even if weak) up to 350 K, which is maximum temperature measured here, and attributed in \cite{5450567,124709} to surface ferromagnetism.

\section{\label{sec:level4}Theoretical framework and computational details}

In the GW approximation (GWA) \cite{PhysRev.139.A796}  the (non-local, dynamical and complex) self-energy $\Sigma$ is given by the convolution in frequency space:
\beq
\Sigma(\bfr_1,\bfr_2,\w) = \frac{i}{2\pi} \int d\w' e^{i\eta\w'} G(\bfr_1,\bfr_2,\w+\w')W(\bfr_1,\bfr_2,\w')
\label{sigma}
\eeq
between the time-ordered one-particle Green's function $G$:
\begin{equation} \label{G}
G(\br_1,\br_2,\omega)
 = \sum_i \frac{\phi_i(\br_1)\phi^*_i(\br_2)}{\omega - \varepsilon_i + i \eta \text{sign}(\epsilon_i-\mu)},
\end{equation}
(where $\mu$ is the Fermi energy) 
and the dynamically screened Coulomb interaction $W$:
\begin{equation}\label{W}
W(\br_1,\br_2,\omega) = \int d\br_3 \epsilon^{-1}(\br_1,\br_3,\omega) v(\br_3-\br_2).
\end{equation}
Here $v$ is the (static) bare Coulomb interaction and $\epsilon^{-1}$ is the inverse dielectric function that describes screening of $v$ through electron-hole and collective plasmon excitations. 
In the GWA $\epsilon^{-1}$ is calculated in the random-phase approximation (RPA).
 Within the quasiparticle self-consistent GW (QSGW) scheme \cite{PhysRevLett.93.126406,PhysRevLett.96.226402} the quasiparticle wavefunctions $\phi_i$ and energies $\varepsilon_i$ are obtained from the (static and Hermitian) potential:
\beq
V = \frac{1}{2} \sum_{ij}\ket{\phi_i} \text{Re} \left[ \Sigma_{ij}(\varepsilon_i) + \Sigma_{ij}(\varepsilon_j) \right] \bra{\phi_j}.
\eeq
At self-consistency the quasiparticle energies $\varepsilon_i$ determine the band structure and
the spectral function is calculated from
\beq
A_{ii}(\w) = \frac{1}{\pi} \frac{\left| \text{Im} \Sigma_{ii}(\w)\right|}
{[\w-\varepsilon_{i}-(\text{Re}\Sigma_{ii}(\w)-V_{ii})]^2+[\text{Im} \Sigma_{ii}(\w)]^2}.
\label{eq:spectral}
\eeq

We have used the QSGW implementation of Ref. \onlinecite{PhysRevB.74.045102} in the Abinit code \cite{abinit}.
We have adopted the MnSi experimental crystal structure from Ref. \onlinecite{crystalstructure}.
We have employed Troullier-Martins norm-conserving pseudopotentials \cite{PhysRevB.43.1993} with Mn $3s$ and $3p$ semicore states treated as valence electrons and  160 Hartree energy cutoff.
We have used a 6$\times$6$\times$6 $\Gamma$-centered k-point grid and 350 (500) bands for the calculation of the screening (self-energy).
The plane-wave size of the dielectric function corresponds to 5.3 Hartree cutoff.
Finally, the frequency integration in Eq. \eqref{sigma} has been performed with a contour-deformation technique using 30  frequencies on the real axis and 6 on the imaginary axis.

\section{\label{sec:level4} Calculated band structure and spectral function}

\begin{figure}[t]
\includegraphics[width=0.5\textwidth]{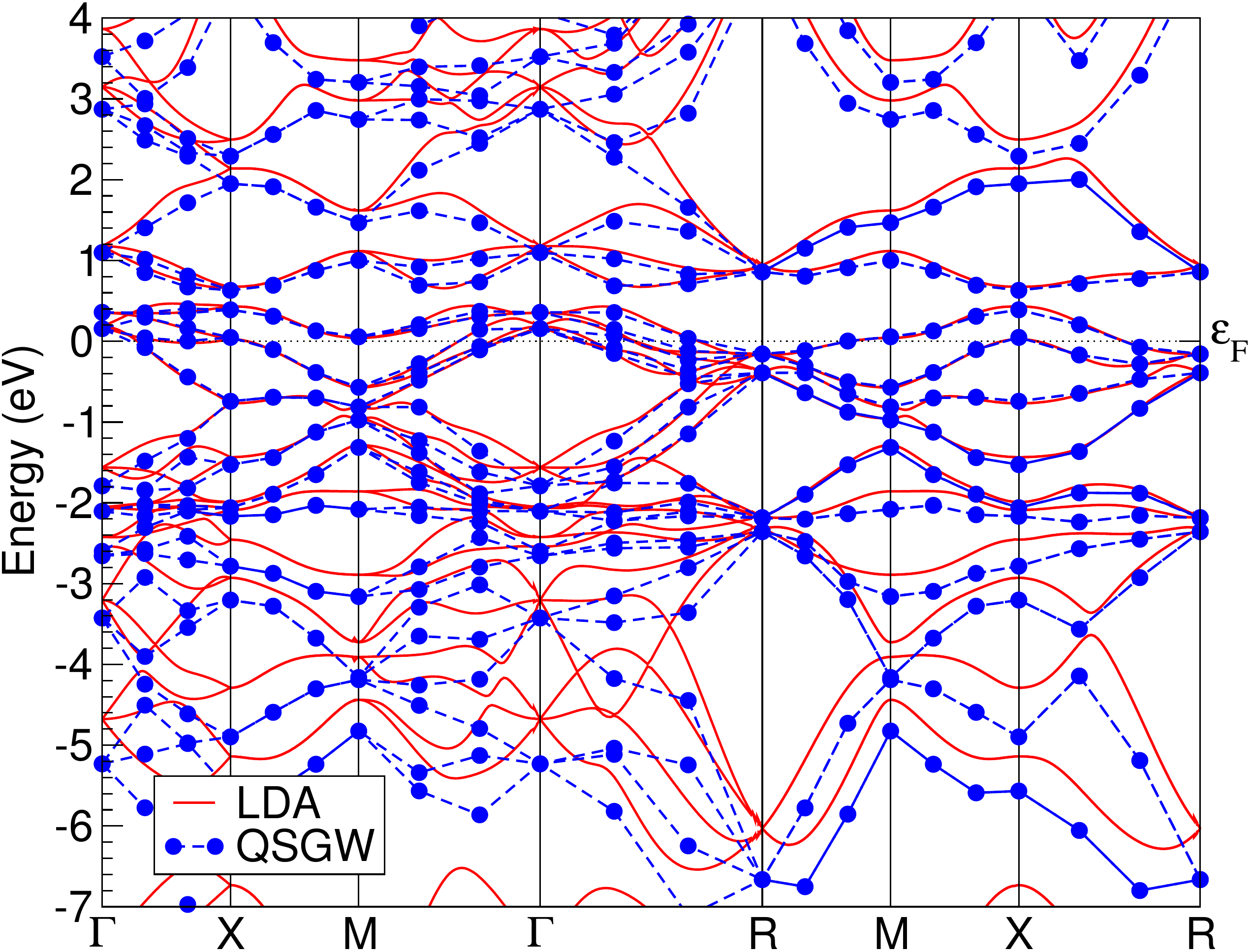}
\caption{\label{bandstructure} Band structure calculated in LDA (red lines) and within QSGW (blue markers; linearly interpolated dashed lines are a guide for the eye).}
\end{figure}

Fig. \ref{bandstructure} shows the band structure of the paramagnetic phase of MnSi calculated in LDA and QSGW.
LDA results are in agreement with those reported by Jeong and Pickett in Ref. \onlinecite{Jeong}.
Since Kohn-Sham eigenvalues in principle are not the electron removal energies measured by ARPES and in order to take into account non-local exchange and correlation effects, we have used a Green's-function approach within the GWA.
Due to the the presence of localized Mn $3d$ states, the accurate QSGW scheme has been employed (beyond standard perturbative corrections to LDA band structure\cite{RevModPhys.74.601}). 

\begin{figure}[t]
\includegraphics[width=0.35\textwidth, angle =-90]{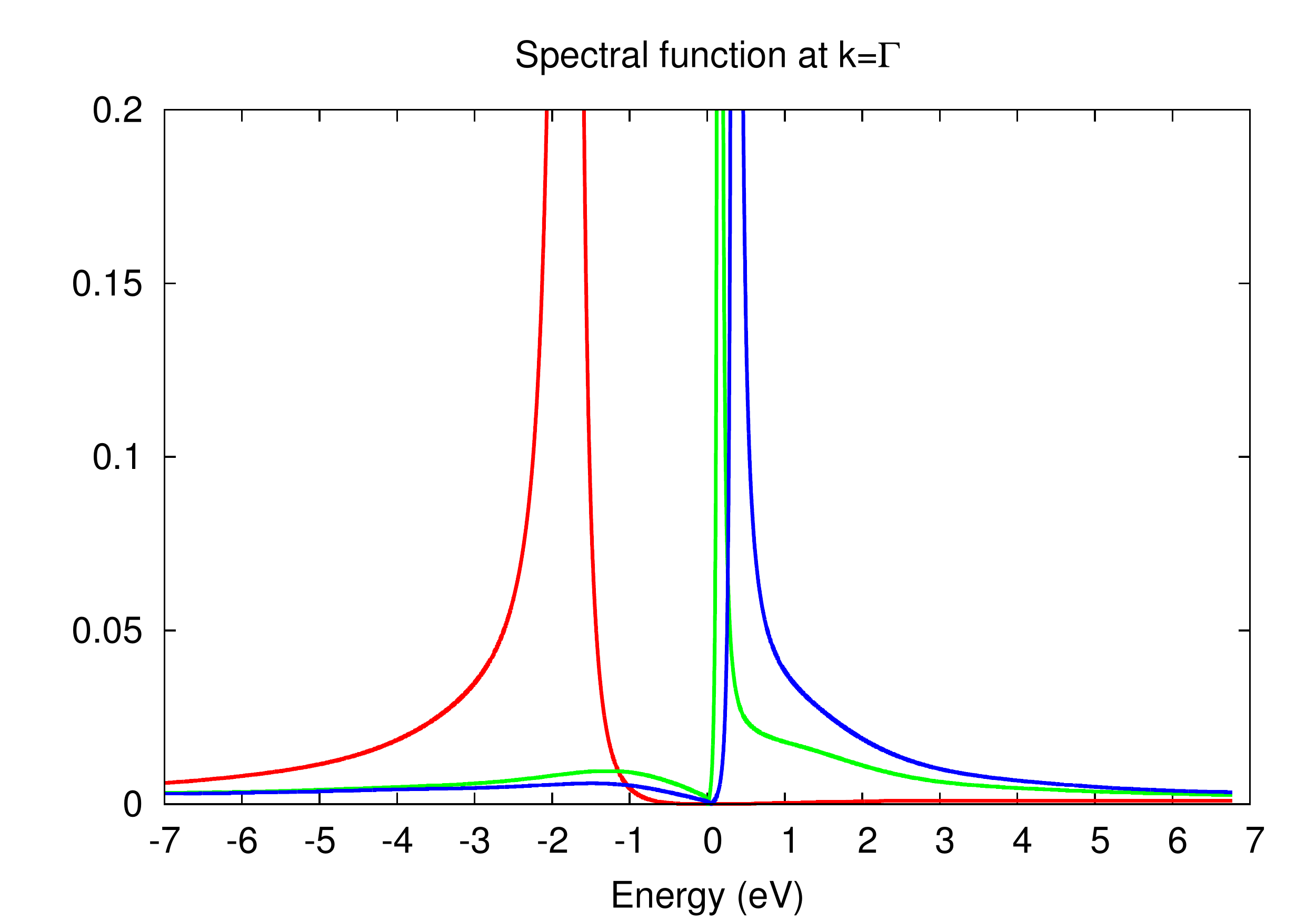}\\
\includegraphics[width=0.35\textwidth, angle =-90]{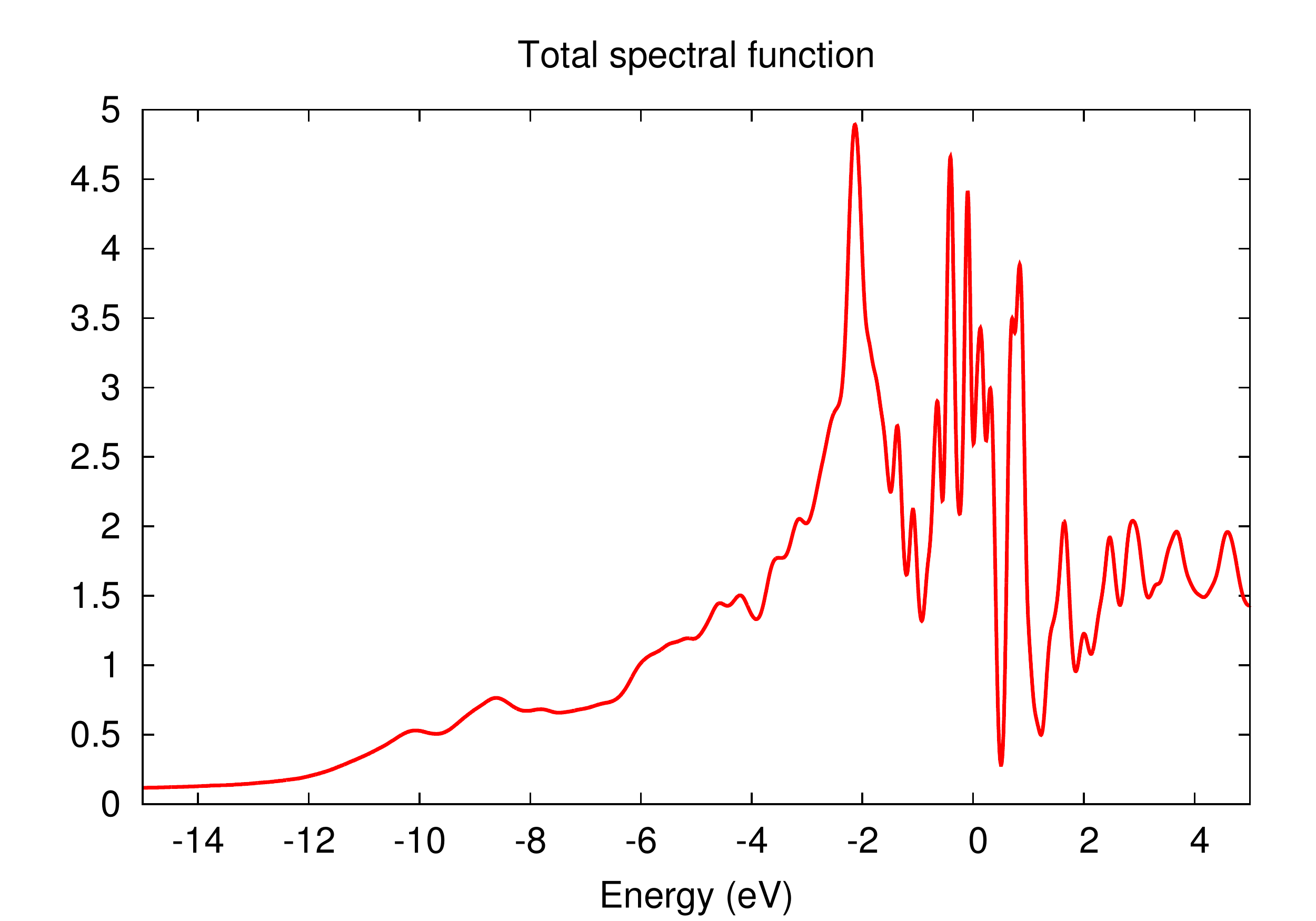}
\caption{\label{spectral} (Top panel) Spectral function for the top-valence and bottom-conduction states at $\Gamma$. (Bottom panel) Total spectral function summed over bands and integrated over the whole Brillouin zone.
The zero of the energy axis is the Fermi level.}
\end{figure}

In the comparison with the LDA band structure, QSGW noticeably differs at energies below $\sim$ -3 eV, increasing the energy separation between Mn $3d$, located around the Fermi energy, and Si $3p$ at higher binding energies. Si $3s$ bands start below -7 eV (not shown).
The same differences between LDA and QSGW appear for the conduction states, where Si $3p$ are present above $\sim$ 1 eV. 
On the contrary, for the Mn $3d$ bands crossing E$_F$ LDA and QSGW are overlapping almost completely. This justifies the use of the LDA for plotting the Fermi surfaces of MnSi.

We have also calculated corrections to the LDA due to spin-orbit coupling and vertex corrections beyond the GWA (in the GW$\Gamma$ approximation with an LDA vertex\cite{PhysRevB.49.8024}) without finding qualitative changes.
Also in the helimagnetic phase (approximated as ferromagnetic in the calculations) the QSGW magnetic moment decreases only by 0.04 $\mu_B/\text{Mn}$ with respect to the LDA\cite{Jeong}, confirming that the theoretical overestimation of the magnetic moment should be fixed by taking into account spin excitations not included in the GWA. Similar differences in the band structure between the LDA and QSGW  as those discussed here for the paramagnetic phase are seen also for the magnetically ordered phase.

The top panel of Fig. \ref{spectral} shows the GW spectral function for the top-valence and bottom-conduction states at $\Gamma$ for paramagnetic MnSi.
The quasiparticle peaks have all a finite width corresponding to a finite lifetime. The incoherent spectral weight appearing below $E_F$ that is associated to the conduction states (i.e. having quasiparticle peaks above E$_F$) is too small to explain the features seen by ARPES at $\sim$~0.5 eV binding energy (see main text).
Finally, the bottom panel  of Fig. \ref{spectral} reports the GW spectral function summed over the valence and conduction states and integrated over the whole Brillouin zone (i.e. it should be compared with angle-integrated photoemission experiments).

\bibliography{Biblio_corrected_Matteo.bib}